\documentclass[sigconf,nonacm]{acmart}

\usepackage{natbib}
\setcitestyle{numbers,square}
\usepackage{amsmath}
\usepackage{cuted}
\usepackage{algorithm}  
\usepackage{algorithmic}  

\usepackage{flushend}
\usepackage{enumerate}
\usepackage{stfloats}

\makeatletter
\newenvironment{breakablealgorithm}
  {
   \begin{center}
     \refstepcounter{algorithm}
     \hrule height.8pt depth0pt \kern2pt
     \renewcommand{\caption}[2][\relax]{
       {\raggedright\textbf{\ALG@name~\thealgorithm} ##2\par}%
       \ifx\relax##1\relax 
         \addcontentsline{loa}{algorithm}{\protect\numberline{\thealgorithm}##2}%
       \else 
         \addcontentsline{loa}{algorithm}{\protect\numberline{\thealgorithm}##1}%
       \fi
       \kern2pt\hrule\kern2pt
     }
  }{
     \kern2pt\hrule\relax
   \end{center}
  }
\makeatother

\renewcommand\footnotetextcopyrightpermission[1]{}

\makeatletter
\renewcommand{\maketag@@@}[1]{\hbox{\m@th\normalsize\normalfont#1}}%
\makeatother

\AtBeginDocument{%
  }

\begin{document}

\raggedbottom

\title{Can Agents Spontaneously Form a Society?}
\subtitle{Introducing a Novel Architecture for Generative Multi-Agent to Elicit Social Emergence}

\author{Hanzhong Zhang}
\affiliation{%
 \institution{Kunming University of Science and Technology}
     \city{Kunming}
 \state{Yunnan Province}
 \country{China}}
\email{armihia@foxmail.com}

\author{Jibin Yin}
\affiliation{%
 \institution{Kunming University of Science and Technology}
     \city{Kunming}
 \state{Yunnan Province}
 \country{China}}
\email{yjblovelh@aliyun.com}

\author{Mulin Jiang}
\affiliation{%
 \institution{Kunming University of Science and Technology}
     \city{Kunming}
 \state{Yunnan Province}
 \country{China}}
\email{237800718@qq.com}

\author{Cong Su}
\affiliation{%
 \institution{Kunming University of Science and Technology}
     \city{Kunming}
 \state{Yunnan Province}
 \country{China}}
\email{787358363@qq.com}

\renewcommand{\shortauthors}{Zhang et al.}
\acmBooktitle{Book title}

\begin{abstract}
Generative agents have demonstrated impressive capabilities in specific tasks, but most of these frameworks focus on independent tasks and lack attention to social interactions. We introduce a generative agent architecture called ITCMA-S, which includes a basic framework for individual agents and a framework called LTRHA that supports social interactions among multi-agents. This architecture enables agents to identify and filter out behaviors that are detrimental to social interactions, guiding them to choose more favorable actions. We designed a sandbox environment to simulate the natural evolution of social relationships among multiple identity-less agents for experimental evaluation. The results showed that ITCMA-S performed well on multiple evaluation indicators, demonstrating its ability to actively explore the environment, recognize new agents, and acquire new information through continuous actions and dialogue. Observations show that as agents establish connections with each other, they spontaneously form cliques with internal hierarchies around a selected leader and organize collective activities.
\end{abstract}
\keywords{Generative agents, Multi agent system, Social interaction, LLM}

\maketitle

\section{Introduction}

Large language models (LLMs) have contributed to significant progress in the field of natural language processing and are widely used in various domains, such as machine translation \cite{xue2024computational}, dialogue generation \cite{cao2023diaggpt}, and content creation \cite{yu2024charpoet}. These models are capable of correctly parsing and generating complex sentence structures and have demonstrated unprecedented capabilities in understanding language. However, LLMs often lack true comprehension and rely more on pattern matching and probabilistic predictions \cite{cheng2024prompt}. It has also been difficult to create systems to achieve human-like systematic generalization \cite{lake2023human}. To overcome these problems, researchers have introduced LLM-based agents. This has allowed the incorporation of external knowledge bases to supplement a model’s knowledge gaps in specific domains \cite{xi2023rise}. They can also be used to decompose a complex task into multiple simpler tasks to achieve hierarchical processing \cite{wei2022chain}.

On this basis, Park et al. \cite{park2023generative} introduced a novel LLM-based agent, namely, a generative agent. This agent simulates trustworthy human behavior. Generative agents have the ability to make multifaceted inferences about an environment, themselves, and other individuals in the environment. They can design daily activity plans based on their own characteristics and experiences and adjust their plans to changes as they occur. When a situation changes, they can flexibly update plans to ensure adaptation to it.

The importance of this progress cannot be ignored. In human – computer interaction, especially in virtual assistants, customer service robots, and even more complex systems, such as self-driving cars and smart homes, the ability of agents to generate believable human behavior is crucial. Generative agents can support more adaptive and flexible interaction processes. This ability not only enhances the system’s responsiveness to dynamic situations but also brings human–computer interaction closer to natural behavioral patterns in interpersonal communication. In addition, by simulating human behavior, generative agents can demonstrate autonomy and sociality in various complex situations, making the interaction process smoother and more intuitive. This feature plays an important role in improving the user experience and increasing the trustworthiness of a system \cite{gao2024taxonomy}.

However, existing generative agent architectures still face many challenges. While traditional agent structures are good at processing and generating behaviors, they are primarily designed for isolated tasks and, thus, mostly lack a focus on sociality. This often makes it difficult for them to model and apply the nuances of social interactions, leading them to focus only on completing tasks and overlook behaviors that promote social connections. This is clearly not conducive to cooperation among multiple agents and may lead to behaviors that are detrimental to the group \cite{ren2024emergence}. In scenarios involving multiple agents, a lack of structured social behavior may lead to disjointed or even chaotic interactions. To truly harness the potential of these agents in domains requiring interaction, there is an urgent need to explore their ability to participate in social interactions, establish relationships, and exhibit emerging social behaviors.

In this paper, we improve upon an existing LLM-based agent architecture (the internal time-consciousness machine based agent [ITCMA] introduced by \cite{zhang2024itcmagenerativeagentbased}) and propose ITCMA-S (the “S” signifies our contribution of social interaction) architecture to enable agents to adapt to multi-agent interaction scenarios. It contains a basic framework for an individual agent and the LTRHA framework for social interaction among multiple agents (the latter is named for its four modules: locale \& topic, resources, habitus, and action; they are described later in the paper). This provides a structured way for agents to identify and filter actions that are not conducive to social interaction, thereby guiding them to choose actions that are more conducive to improving the social atmosphere of a scene. It also allows them to form new relationships during interactions with other agents, while remembering the history of these interactions. The formation of these dynamic relationships occurs naturally and is not pre-programmed. This means that the agent has the ability to adapt to its social environment.

To understand whether and how ITCMA-S leads to the emergence of sociality among generative agents, we established an evaluation environment called IrollanValley. This is a sandbox world designed to test and allow the observation of generative agent behavior and social interactions. It contains six characters, eight areas, and six operational primitives. Each area has unique furniture and other items, and an agent can interact with the various characters and objects and move around in the areas. Based on this environment, we modeled the natural evolution of social relationships among multiple agents without identities in the environment. The human evaluation results showed that the agents had spontaneously developed good social skills on each evaluation indicator and were able to actively explore the environment, meet new agents, and acquire new information through continuous actions and conversations. By observing the environment, we found that, in the process of establishing connections among each other, the agents spontaneously formed a clique with internal hierarchies around an elected leader and organized collective activities under the leader’s guidance. Agents not included in this clique were mostly in isolated states. Even when they were in the same room as other agents, they interacted less and focused mainly on their personal activities.

In summary, this paper makes the following contributions: 

\begin{itemize}
\item We propose a generative multi-agent structure, ITCMA-S, which includes the structure of an individual generative agent and a multi-agent social collaboration framework, LTRHA.

\item We established an environment for multi-agent social evaluation, IrollanValley, which consists of six roles, eight areas, and six operational primitives designed to assess the utility and sociality of multiple agents.

\item We conducted validation and ablation studies in IrollanValley through human evaluation, and the results showed that ITCMA-S performed excellently across multiple indicators.

\item Furthermore, we discovered that agents can actively explore the environment, cooperate with other agents through a division of labor, and spontaneously form small groups and leadership structures in complex social settings.

\end{itemize}

Section 2 of this paper introduces related works; Section 3 introduces the improvement of the individual agent structure of ITCMA-S on an existing generative agent architecture and includes an explanation of how it generates and infers action through perception, memory, and emotion-driven mechanisms; Section 4 introduces a multi-agent interaction framework in ITCMA-S, namely LTRHA; Section 5 introduces the experimental evaluation of the sociality of ITCMA-S and discusses the process and results of the formation of the agents’ social relationships; and Section 6 summarizes the main findings of the study and suggests future research potential.

\section{Related Work}

\subsection{LLM-based agent}

Park et al. \cite{park2023generative} introduced a generative agent based on an LLM and demonstrated that it can generate trustworthy individual behavior and sudden group behavior in simulations. For example, a generative agent will turn off a stove when it sees its breakfast burning and will stop to chat when it encounters other agents with whom it wants to talk. Based on this research, numerous LLM-based agents have emerged. Zhang et al. \cite{zhang2023appagentmultimodalagentssmartphone} proposed AppAgent, which constructs an agent to operate any smartphone application. Hong et al. \cite{hong2023cogagentvisuallanguagemodel} proposed an 18-billion-parameter visual language model (VLM) named CogAgent, which specializes in GUI understanding and navigation. Vezhnevets et al. \cite{vezhnevets2023generativeagentbasedmodelingactions} provided the Concordia library to simulate agent interactions in physical, social, and digital spaces. Among these, a special agent called the game master (GM) is responsible for simulating the environment of agent interactions. Agents take action by describing what they want to do in natural language, and the GM then translates their actions into appropriate implementations.

\subsection{Research on the Structure of Generative Agents}

The original structure of generative agents, as described by Park et al. \cite{park2023generative}, mainly consisted of three parts: memory flow, reflection, and planning. The concept of a chain of thought (CoT) was particularly important for improvements to the planning module \cite{wei2022chain}. CoT refers to the ability of an LLM to think and reason gradually through a series of steps or iterations, reflecting human cognitive processes. Traditional language models generate responses without clear intermediate steps, which can lead to suboptimal answers, especially in complex inference scenarios. CoT overcomes these limitations by introducing intermediate steps to enable language models to reason, thereby enhancing the model’s ability to solve problems. Mondal et al. \cite{mondal2024kam} suggested using knowledge graphs to enhance multiple patterns to help models solve complex problems, thereby triggering CoT functionality. Their proposed method, knowledge augmented multimodal (KAM)-CoT, decouples the inference process into two consecutive stages. In the first stage, practical reasons are provided, and in the second stage, the generated reasons are used as additional input to provide answers.

As an improvement to the memory module, Liu et al. \cite{liu2024llmconversationalagentmemory} proposed the reasoning and acting through scratchpad and examples (RAISE) architecture, which was specifically designed to enhance the functionality of conversational agents. It combines a dual-component memory system, similar to the short-term and long-term memory functions of the human brain. Toy et al. \cite{toy2024metacognitionneedusingintrospection} proposed a metacognitive module by improving the reflection module in generative agents, allowing agents to broadly consider their situations to create alternative strategies and improve their performance.

To improve the overall structure of generative agents, Wu et al. \cite{wu2023plan} explored an alternative mechanism that utilizes prior knowledge encoded in an LLM without affecting the trainability of agents. Their proposed three-step PET framework: Plan, Eliminate, and Track. Lin et al. \cite{lin2024swiftsage} proposed the method SwiftSage, which can achieve both fast and slow thinking in complex interactive reasoning tasks. It effectively blends the advantages of behavioral cloning and LLM.

\subsection{Research on the Sociality of Generative Multi-Agents}

By interacting with each other, generative agents can exchange information, form new relationships, and coordinate activities. These social behaviors occur naturally rather than being pre-programmed. Over time, agents form new relationships and remember their interactions with other agents \cite{park2023generative}. Generative Multi-Agents can simulate complex human systems. Wang et al. \cite{wang2024largelanguagemodelsneed} designed a new framework called the mosaic expert observation wall (MEOW). In MEOW, real game data are processed by expert models trained on simulated data and converted into natural language prompts to assist LLM inference.

However, while an LLM can capture social norms, there is also research that suggests that an LLM cannot adequately understand social norms, especially culture-specific social norms. This defect may lead to conflicts between generative agents, especially when their underlying LLMs are trained on text corpora from different cultural backgrounds \cite{hämmerl2022multilinguallanguagemodelscapture, ramezani2023knowledge}. Ren et al. \cite{ren2024emergence} proposed the specification architecture CRSEC (named for its four modules: Creation \& Representation, Spreading, Evaluation, and Compliance) for generative multi-agent systems to resolve this problem. Building on this foundation, these researchers examined the emergence of social norms in generative multi-agent systems. Ghaffarzadegan et al. \cite{ghaffarzadegan2024generative} provided a new approach to developing models that reduces reliance on assumptions about human decision-making and utilizes the vast amount of data in LLMs to capture human behavior and decisions. By utilizing the extensive dataset in the LLM, their generative agent-based model helps represent human decisions in computational models. Motwani et al. \cite{motwani2024secretcollusiongenerativeai} pointed out that groups of generative agents can use communication channels in ways that are unexpected to their developers. When sharing data, stealthy collusion can help agents coordinate unwanted behavior on a larger scale, and GPT-4 has demonstrated unparalleled capabilities in this regard.

\section{Generative Agent Framework}

\subsection{The Internal Time-Consciousness Machine Based Agent}

Zhang, Yin, et al. \cite{zhang2024itcmagenerativeagentbased} introduced the internal time-consciousness machine (ITCM), which is a computational consciousness structure. The ITCM supports agents in taking action and making inferences in an open world. It can help generative agents become more flexible and intelligent when handling complex tasks; this improves their interpretability and makes their actions easier to understand and predict. On this basis, they proposed ITCM-based agent (ITCMA). As a generative model, ITCMA considers both the reasoning ability of agents and the interaction between agents and the environment to compensate for the shortcomings of LLMs in accomplishing specific tasks.

ITCMA can be simply explained as a framework: it uses a spherical coordinate space called a “phenomenal field” as one time frame (1\emph{f}) of perception representing a certain moment, and constructs a time-continuous consciousness channel $C^t$ of equivalent working memory through a field string composed of retention (perception of the past) $Re^t$ and primal impression (perception of the present) $PI^t$ at time $t$ in units of \emph{f}. The $C^t$ can be combined with the long-term memory $AM$ of the agent activated by perception, and the possible changes of protention (prediction of future perception) $Pro^t$ when an action is to be taken that can be deduced through the time-series forecasting model (TSFM). After that, it blends into a natural language format together with the intrinsic motivation $d^t$, which includes the agent’s emotional state (including pleasure, arousal, and dominance) and is provided to the LLM to generate the output of the action. Its structure is shown in Figure 1.

\begin{figure}[h]
  \centering
  \includegraphics[width=\linewidth]{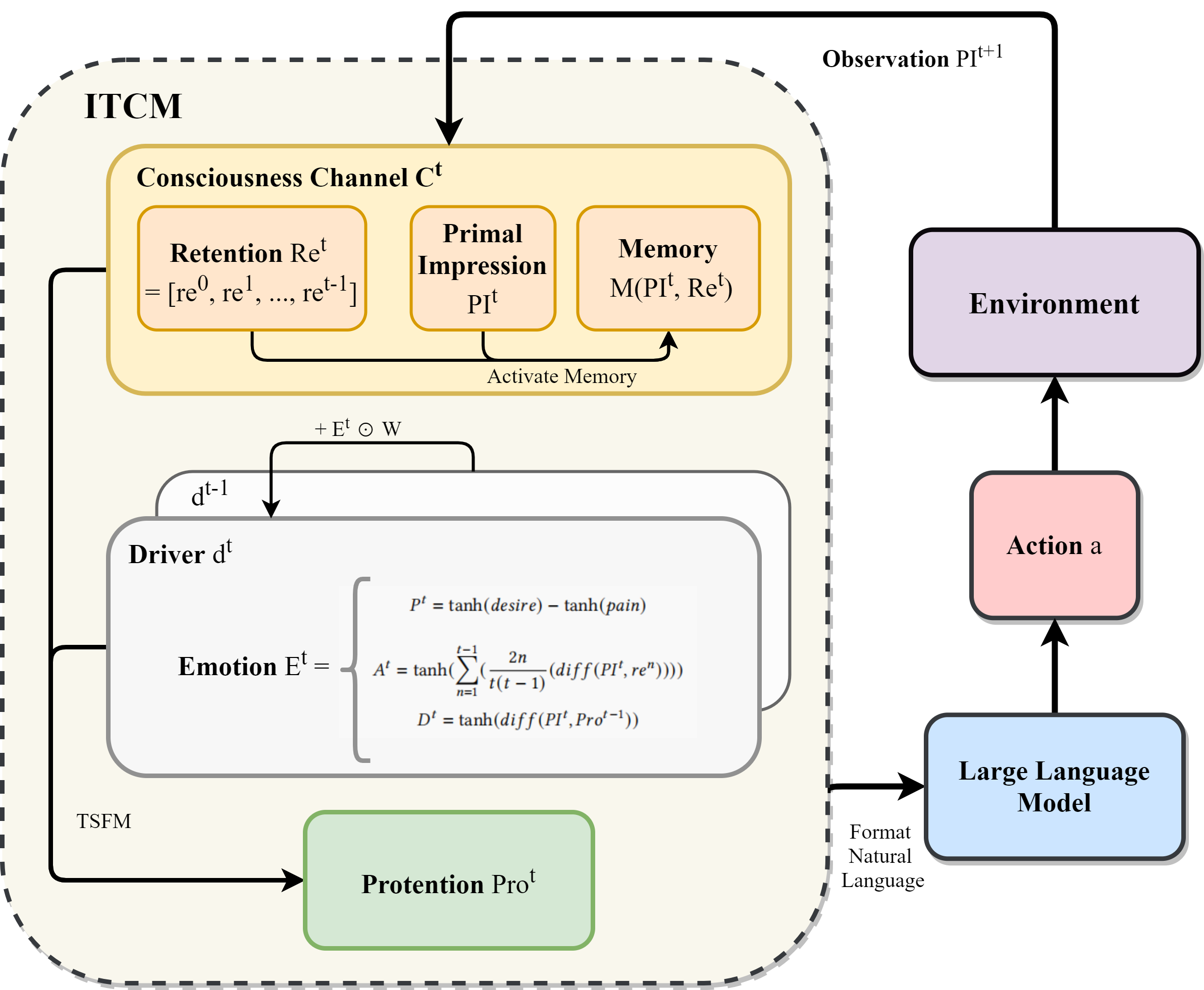}
  \caption{Structure of ITCMA. ITCMA’s main structure, the ITCM, contains the consciousness channel, driver, and protention. In the consciousness channel, retention and primal impression are used to activate memory, and the driver consists of the dimensions of emotion and the driver from the previous moment. These two are used to predict protention through the TSFM. The content of the ITCM will be converted into formalized natural language, which will be delivered to the LLM and will eventually cause it to act.}
\end{figure}

Through this mechanism, an agent perceives inputs from the environment and takes action as output to the environment. The “brain” of an agent is not equivalent to an LLM here but is replaced by a complete ITCMA structure including the agent’s memory. The LLM is only used as a tool. Experiments with ITCMA have demonstrated the effectiveness and generalization of this architecture. The trained agents exceeded the state of the art (SOTA) in the environment, and even completely untrained agents can start a task very quickly after exploring the environment and achieve good results.

However, the structure of ITCMA still has shortcomings, including slow processing speed, excessive token consumption, and slow entry into the task context when untrained. To better test our assumptions about its performance, we first made certain improvements to the structure of ITCMA, resulting in a generative multi-agent model for social interaction, namely ITCMA-S. It includes a structure for an individual agent and a structure for multi-agent interaction. The structure for an individual agent improved the modules of memory, motivation, and action space reduction based on ITCMA and are elaborated in the remaining parts of this section, while the structure for multi-agent interaction are discussed in Section 4.

\subsection{Memory and Imagination}

\subsubsection{Memories Blended into the Present}

In ITCMA, after a memory is awakened to the present moment, it is juxtaposed with the retention and primal impression in the consciousness channel. However, the theory of creature consciousness suggests that phenomenal consciousness requires the blending of a “phenomenal field” mechanism that may originate in the thalamus and neural inputs from different cortical areas responsible for processing memory-related information \cite{bayne2007conscious}. It is obvious that, for humans, the awakening of memory is not simply juxtaposition but blending with present consciousness. Conceptual blending is a cognitive activity that combines information from different contexts \cite{fauconnier2024way}. Its main process is composition, which is the process of projecting input spaces (two different fields) into the blended space. Blending can combine elements from the input spaces to provide relationships that do not exist within a single input space.

\begin{figure}[h]
  \centering
  \includegraphics[width=\linewidth]{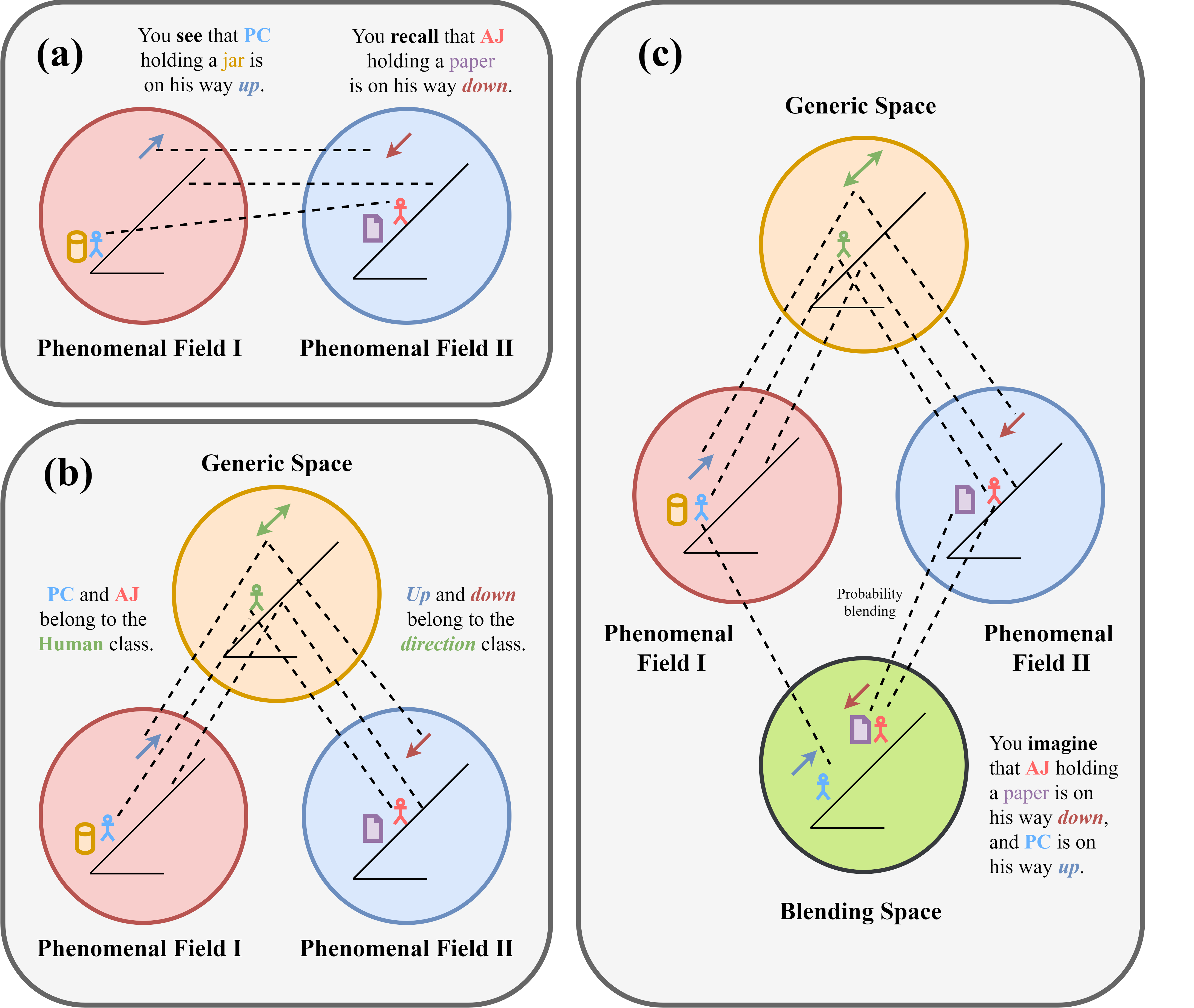}
  \caption{The Conceptual Blending Process of the Phenomenal Field. (a) Two phenomenal fields are matched to create cross-space mapping; (b) a generic space is established for the matched objects in the two phenomenal fields; and (c) with the help of the generic space, the components and structures in the two phenomenal fields selectively enter the blended space.}
\end{figure}

Therefore, based on conceptual blending theory, we hypothesize that when a memory of ITCMA-S enters the current consciousness channel, its phenomenal fields of observation and recollection are blended to obtain the imagination of this moment, and thus the material of the consciousness channel is obtained. As shown in Figure 2, this process satisfies the following steps: In the first step, Figure 2(a), there is local matching between the phenomenal fields; that is, the equivalent component connections are generated by the matching. Once the match between the two fields is created, it is said that there is cross-space mapping between them. In the second step, Figure 2(b), which is the blending process, the matching structure of the two phenomenal fields is utilized to establish the generic space (which can be roughly understood as the common “belonging” class containing the instances of the elements of the phenomenal fields). In the third step, Figure 2(c), via the generic space, the two phenomenal fields are projected into a new space: the blended space. After this, components and structures in the phenomenal field selectively enter the blended space, forming structures that are, to some extent, distinct from the original phenomenal field.

We stipulate that for completely identical objects, they are placed in the blended space after blending takes place (i.e., taking the average); objects that match in the generic space (with similarity exceeding a threshold) are each placed in the blended space; and objects that do not match have a certain probability of being placed in the blended space. Therefore, for the two phenomenal fields $f^x$ and $f^y$, the blending process $Blend(f^x, f^y)$ follows Algorithm 1:

\begin{footnotesize}
\begin{equation}
\left.
\mathrm{f}^x=
\left[
    \begin{matrix}
        \mathrm{f}^x{}_1\\
        \vdots\\
        \mathrm{f}^x{}_a
    \end{matrix}
\right.\right]
=
\left[
    \begin{matrix}
        N^x{}_1&pos^x{}_1\\
        \vdots&\vdots\\
        N^x{}_a&pos^x{}_a
    \end{matrix}
\right]
=
\left[
    \begin{matrix}
        N^x{}_1^1&\cdots&N^x{}_1^n&\theta^x{}_1&\varphi^x{}_1&\gamma^x{}_1\\
        \vdots&\ddots&\vdots&\vdots&\vdots&\vdots\\
        N^x{}_a^1&\cdots&N^x{}_a^n&\theta^x{}_a&\varphi^x{}_a&\gamma^x{}_a
    \end{matrix}
\right]
\end{equation}
\end{footnotesize}

\begin{footnotesize}
\begin{equation}
\left.\mathrm{f}^y=\left[\begin{matrix}\mathrm{f}^y{}_1\\\vdots\\\mathrm{f}^y{}_b\end{matrix}\right.\right]=\left[\begin{matrix}N^y{}_1&pos^y{}_1\\\vdots&\vdots\\N^y{}_b&pos^y{}_b\end{matrix}\right]=\left[\begin{matrix}N^y{}_1^1&\cdots&N^y{}_1^n&\theta^y{}_1&\varphi^y{}_1&\gamma^y{}_1\\\vdots&\ddots&\vdots&\vdots&\vdots&\vdots\\N^y{}_b^1&\cdots&N^y{}_b^n&\theta^y{}_a&\varphi^y{}_b&\gamma^y{}_b\end{matrix}\right]
\end{equation}
\end{footnotesize}

\begin{equation}
\begin{split}
&SphericalSim(A, B)=\\
&1-\frac{1}{3}(\omega_{\gamma}\tanh(\left| \gamma_{A}-\gamma_{B} \right|)+\omega_{\theta}\frac{\left| \theta_{A}-\theta_{B} \right|}{\pi}+\omega_{\varphi}\frac{\left| \varphi_{A}-\varphi_{B} \right|}{2\pi})
\end{split}
\end{equation}

\begin{equation}
\begin{split}
&FieldSim(f^x, f^y)=\\
&\cfrac{\sum_{i=1}^a\omega_{N}Cosin({N^x}_{i}, {N^y}_{j})+\omega_{pos}SphericalSim({pos^x}_{i}, {pos^y}_{j})}{Max(a,b)}
\end{split}
\end{equation}

\begin{breakablealgorithm}
\caption{Conceptual Blending Algorithm of Phenomenal Field $Blend(f^x, f^y)$}

\begin{algorithmic}[1]
    \REQUIRE Initial Fields $f^x$ and $f^y$
    \ENSURE Blended Field $f^z$
    
    \STATE Initialize threshold of similarity degree $T$
    \STATE Initialize blended probability $r$
    
    \FOR{$i \, from \, 1 \, to \, a$}
        \FOR{$j \, from \, 1 \, to \, b$}
            \STATE $s \leftarrow FieldSim(f^x_{i}, f^y_{j})$
            \IF{$s<T$}
                \STATE \textbf{Continue}
            \ELSIF{$s=1$}
                \STATE $f^z_{k} \leftarrow Average(f^x_{i}, f^y_{j})$
                \IF{$f^z_{k}$ not in $f^z$}
                    \STATE Add $f^z_{k}$ as a row to $f^z$
                \ENDIF
                
                \STATE \textbf{Break}
            \ELSE
                \IF{$f^x_{i}$ not in $f^z$}
                    \STATE Add $f^x_{i}$ as a row to $f^z$
                \ENDIF
                
                \IF{$f^y_{j}$ not in $f^z$}
                    \STATE Add $f^y_{j}$ as a row to $f^z$
                \ENDIF
                
                \STATE \textbf{Break}
            \ENDIF
            
        \ENDFOR
        
    \ENDFOR
    \STATE There is a probability that rows in $f^x$ and $f^y$ that are not in $f^z$ will each have rate $r$ added as a row in $f^z$.
    
    \STATE \RETURN $f^z$

\end{algorithmic}
\end{breakablealgorithm}

Among them, $Cosin(A, B)$ represents the cosine similarity between $A$ and $B$. For the retention $Re^t$ and primal impression $PI^t$, the acquisition of protention $Pro^t$ can be corrected as follows:

\begin{equation}
\mathrm{imagine^t=Blend(PI^t,M(PI^t,Re^t))}
\end{equation}

\begin{equation}
\mathrm{C^t=[~Re^t,PI^t,M(PI^t,Re^t)~]}
\end{equation}

\begin{equation}
\mathrm{Pro^t=TSFM(C^t,imagine^t,d^t),}
\end{equation}
where $C^t$ is the content of the consciousness channel at time $t$, TSFM is the selected time-series forecasting model, and $M$ is the process of triggering inventory memory through $PI^t$ and $Re^t$.

\subsubsection{Memory Storage and Compression}

One of the reasons for the slow processing speed of ITCMA is the memory activation algorithm it uses. Its improved Levenshtein distance method provides better memory query results. However, it consumes more time than the simple cosine similarity algorithm. For this reason, it was necessary to improve it.

One way to improve the speed of memory activation is to reduce the number of memories that the agent needs to query. Due to the learning of new memories, the retention and association of existing memories will be affected \cite{anderson2020cognitive}. One way to address this is to modify, split, and recombine memories over time, that is, to compress memories \cite{damasio2012self}. This compression can affect the accuracy of memory, but its impact on recall is not as severe. In fact, people do not use precise memories when making decisions \cite{bransford1972sentence}. The compression process of memory constantly generates new meanings \cite{fauconnier2024way}, and making decisions based on such memories is sometimes seen as intelligent inference, which may be a source of the generalization of human intelligence \cite{reder1982plausibility}.

Therefore, by compressing old memories and blending them when they are recalled to the present moment, we enable ITCMA-S to reduce the total number of memories while preserving the recall effect, thereby increasing the speed of the memory search. In addition, based on the mood congruence effect, we set the weight of the memory index to the emotional intensity accompanying the agent at the time of memory occurrence \cite{eich1994mood, levine1997beyond, damasio2012self}. Specifically, to compress a segment of memory, we need to select the key frame $f^{key}$ that has the strongest arousal in that segment of memory. The process $MBlend_{n}(\text{f}^\text{\:key},\text{Memory}^{1,\text{n}})$ of compressing a segment of memory $\text{Memory}^{1,\text{n}}$ with a length of $n$ by $f^{key}$ is expressed as follows:

\begin{equation}
\mathrm M\mathrm e\mathrm m\mathrm o\mathrm r\mathrm y^{1,\mathrm n}=[\mathrm f_1,\mathrm f_2,\mathrm f_3\dots,\mathrm f_n]
\end{equation}

\begin{equation} 
\begin{split}
\hspace{-12mm}
&MBlend_n(\text{f}^\text{\:key},\text{Memory}^{1,\text{n}})=\\
&\begin{cases}\qquad\qquad\qquad\qquad\text{f}^\text{\:key}&if n=0.\\\text{Blend}(MBlend_{n-1}(\text{f}^\text{\:key},\text{Memory}^{1,\text{n}-1}),\text{f}_\text{n})&otherwise.\end{cases}
\end{split}
\end{equation} 

After compression is complete, $MBlend_{n}(\text{f}^\text{\:key},\text{Memory}^{1,\text{n}})$ replaces the original $\text{Memory}^{1,\text{n}}$ position in the long-term memory base to reduce the total number of memories.

\subsection{Emotion and Motivation}

Emotions can help with decision-making, not just interfere with it, as most people believe \cite{damasio2018}. For example, in decision-making, emotions can highlight the importance of a certain premise, thereby making a decision tend toward that premise. This idea coincides with the practice in ITCMA of allowing LLMs to choose which action to perform by deducing the protention $Pro^t$ of each action in the action space. However, although ITCMA provides LLMs with an internal driving force $d^t$ that includes emotions for decision-making, it only adds the weighted sum of the three-dimensional PAD emotions at this moment to the internal driving force $d^{t-1}$ at the previous moment. Indeed, agents tend to push pleasure and dominance to have the highest values possible \cite{skinner1996guide, diener2000subjective}, but they also generally tend to keep emotional arousal at a stable value \cite{gross1998emerging}. In ITCMA-S, the arousal $A^t$-based mechanism of passive attention is quantified as the degree of change between the elements in retention and the elements in the current primary impression. The dominance $D^t$ is quantified as the difference between the protention $Pro^{t-1}$ at the previous moment and the primal impression $PI^t$ at this moment. The pleasure $P^t$ is quantified as the degree of satisfaction with the agent’s $desire$ and the degree of avoidance of $pain$. The expression $\omega_{P}+\omega_{A}+\omega_{D}=1$ contains the dynamic weights of the emotions. Therefore, for the emotional dimensions $P^t, A^t, D^t \in (-1, 1)$ at time $t$, the emotional values should be considered as follows:

\begin{equation}
\mathrm{P^t}=\tanh(desire)-\tanh(pain)
\end{equation}

\begin{equation}
\mathrm A^t=\tanh(\sum_{n=1}^{t-1}(\frac{2n}{t(t-1)}( diff(\mathrm P\mathrm I^t,\mathrm r\mathrm e^\mathrm n) ) ))
\end{equation}

\begin{equation}
\mathrm D^t=\tanh( diff(\mathrm P\mathrm I^t,\mathrm P\mathrm r\mathrm o^{\mathrm t-1}) )
\end{equation}

\begin{equation}
\mathrm{d_{bias}}^{t}=\omega_{\mathrm{P}}\mathrm{P^{t}}+\omega_{\mathrm{D}}\mathrm{D^{t}}+\omega_{\mathrm{A}}(1-\mid\mathrm{A^{t}}-\mathrm{A^{t-1}}\mid)
\end{equation}

Among them, the calculation of $desire$ and $pain \in [0, \infty)$ are defined according to specific situations. For example, in reinforcement learning tasks, $desire$ can be defined as reward, while $pain$ can be defined as punishment. However, to avoid the situation of local optima (where agents enter a scenario in which they can continuously obtain $desire$ and reduce $pain$ without taking any further actions), we introduce a demand dimension $Ne$ for them.

The demand motivation model \cite{maslow2013motivation,zhang2022needs} states that when an agent falls into a local optimum and its basic needs are satisfied, these needs are no longer important, and the agent moves on to more advanced needs. As a result, the agent needs to pursue advanced needs to gain new $desire$, and the $desire$ then gained from satisfying basic needs falls while the $pain$gained rises. This can be approximated as dynamic rewards and punishments. For ${d_{bias}}^{t} \in (-1, 1)$, these are as follows:

\begin{equation}
\mathrm{Ne}=\frac{\mid\mathrm{d_{bias}}^{t}-\mathrm{d_{bias}}^{t-1}\mid}{2}
\end{equation}

\begin{equation}
\mathrm{d^t=d^{t-1}+Ned_{bias}^ {\quad \:t}}
\end{equation}

\subsection{Reduction of Action Space}

The PET framework created an elimination module for AlfWorld using pre-trained Q\&A models to filter out containers and objects unrelated to the current task based on common sense about the task \cite{wu2023plan}. Through this step, the time required for an untrained agent to enter the task context can be effectively reduced. Similarly, ITCMA-S also uses an LLM $Elim$ to reduce the number of action spaces in a zero-shot manner. Specifically, for the target $G$ of the agent, we create a prompt in $Desc=$ “Your task is to: $G$. The actions you can take are: $AS$. The $a_i$ will be relevant?” format for the executable action $a_i \in AS$ in the action space $AS$. $Elim$ will output the confidence score for the action $a_i$:

\begin{equation}
\mu_{\mathrm{a_i}}=Elim(Desc,G,AS,\mathrm{a_i})
\end{equation}

Among them, $\mu_{\mathrm{a_i}} \in [1, 5]$. When $\mu_{\mathrm{a_i}}$ is less than the threshold, $a_i$ will be removed from the action space. The elimination process is shown in Figure 3.

\begin{figure}[h]
  \centering
  \includegraphics[width=\linewidth]{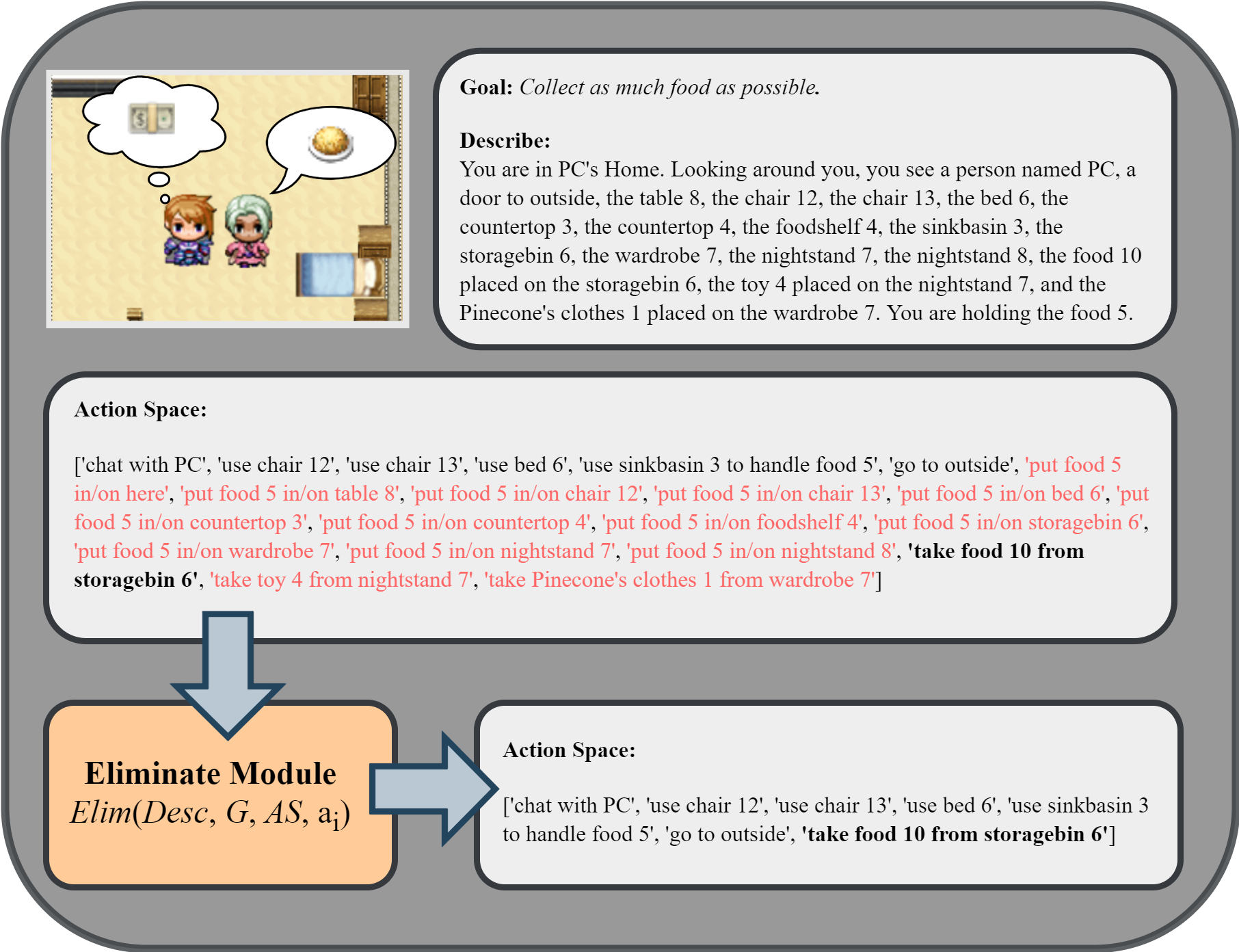}
  \caption{The Process of Reducing the Action Space. Due to the high degree of freedom in the scene, there are too many available actions. Among them, the red action is unrelated to the current goal, so it should be eliminated without the need for further protention calculation.}
\end{figure}

\section{Social Interaction Framework}

Zhang, Duan, et al. \cite{zhang2024tribaltheatermodelsocial} proposed a social regulation model for the dynamic adaptation of users in virtual interactive environments, namely the tribal theater model (TTM), to address the core issue of “enhancing user interaction freedom.” This model emphasizes the subjectivity of interactive users. In this section, based on the TTM and field theory \cite{bourdieu2015invitation}, we present a multi-agent social interaction architecture, the LTRHA, for ITCMA-S, which consists of four modules: locale \& topic, resources, habitus, and action. We aimed to design an interaction architecture for generative agents to promote the emergence of spontaneous social interactions within their societies. Specifically, there is no preset identity in the LTRHA environment. Every agent has certain resources. The environment will provide basic action options, such as using objects and communicating with other agents. The probability of successfully executing these actions will depend on the number of resources and will be handed over to the matrix model for processing after execution. Along with the actions input to the model, there is a vector consisting of the current resource structure of the environment and a vector of topics that have been quantified according to the analysis of the emotions of each agent. Based on the input, the matrix model will adjust the resource structure of the environment and output it as a vector to change the resource structure. The specific process is shown in Figure 4. We provide a detailed introduction to each part of the LTRHA framework in the sections that follow.

\begin{figure}[h]
  \centering
  \includegraphics[width=\linewidth]{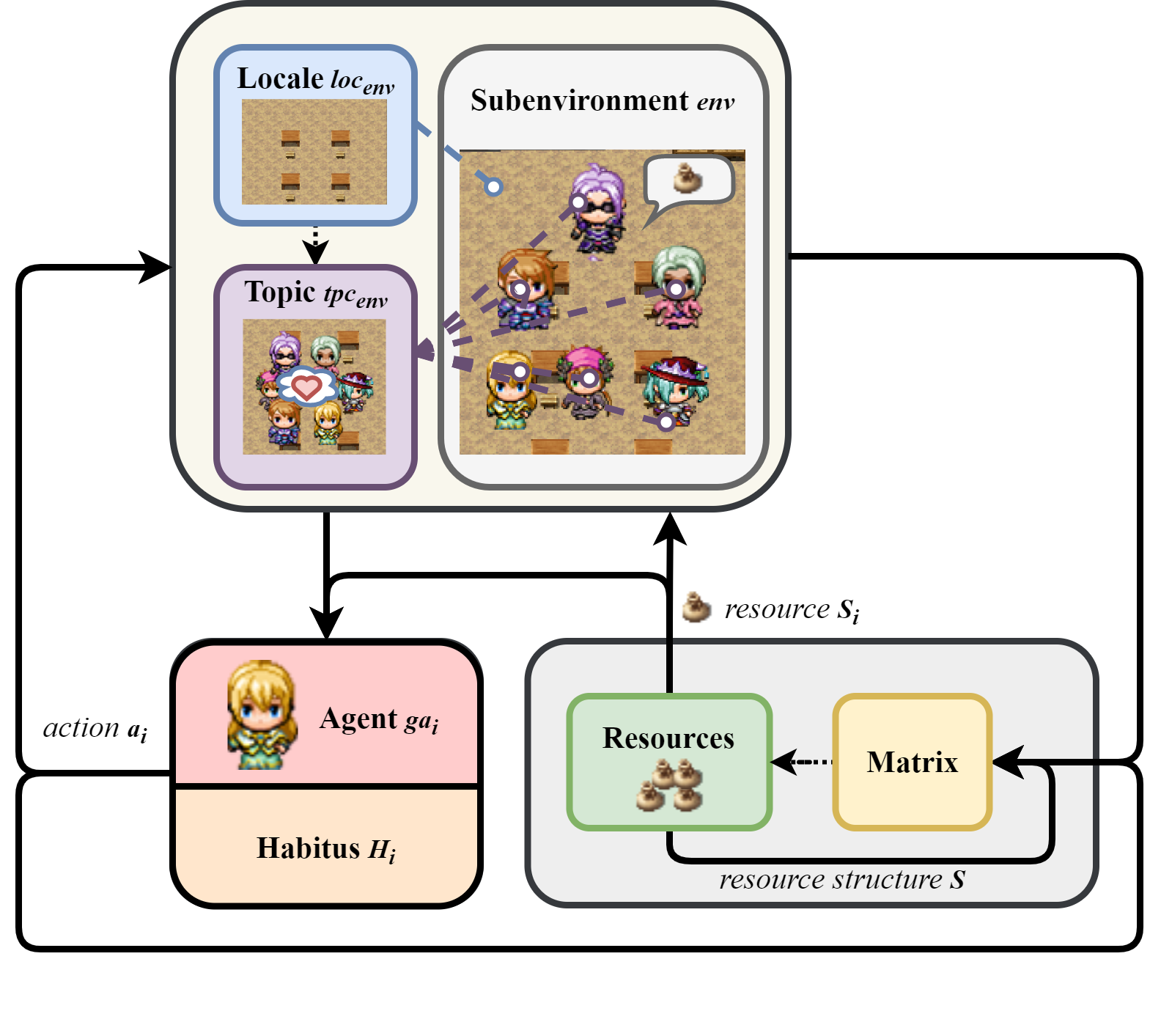}
  \caption{Execution Process of the LTRHA. An agent executes actions with probabilities related to the number of resources and then hands over the actions, the current resource structure, and quantified topics of the environment to the matrix model for processing. The matrix model adjusts the resource structure of the environment based on input.}
\end{figure}

\subsection{Locale and Topic}

In the TTM, the field is decomposed into two parts: the tribe and the atmosphere \cite{zhang2024tribaltheatermodelsocial}. To distinguish the field in ITCMA-S from the field in ITCMA, we refer to the parts as the locale and topic. The locale and topic modules can be understood together as an interactive space. In human society, such interactive spaces typically include locales as physical spaces and topics as mental factors \cite{maffesoli2022time,norberg-schulz2010genius}. For example, in a speech setting, the stage and audience seats are part of the locale, and the passion aroused by the speaker in the audience is a topic.

We define a sub-environment $env$ of the overall environment, which includes a space and $n$ agents occupying that space. We make the spatial area and its contained objects $loc_{env}$ as the locale, and the emotion synthesis function $tpc_{env}$ of these agents is a topic. Among them, we define agent $ga_i \in GA$ with emotional dimensions $P_i, A_i, D_i \in (-1, 1)$. The emotion synthesis function is described by the following equation:

\begin{equation}
tpc_{env}=\frac{\sum_{i=1}^{n}P_{i}\frac{(A_{i}+1)}{2}}{n}
\end{equation}

As a result, the topic value of a sub-environment will be in the interval $[-1,1]$, where less than zero is a negative atmosphere and more than zero is a positive atmosphere. In $env$, any agent can change the perception of other agents by affecting the objects in $loc_{env}$, thereby affecting $tpc_{env}$.

\subsection{Resources}

The execution of actions by agents requires a certain amount of resources, just as implementing decisions in human society requires a certain cost. Resources are allocated to the agents in the sub-environments based on the actions of the agents by a model called the “matrix.”

\subsubsection{Competition for Limited Resources by Agents}

The key to interactive regulation is the resources possessed by an agent \cite{grenfell2012pierre, bourdieu2015invitation}. Virtual resources, such as the cultural level and social status, are considered interactive resources. We believe that the agents’ environment is a space for competing resources. Changing the distribution and relative weights of resources is equivalent to changing the structure of the environment. A resource is both a weapon and an object of contention, enabling its owner to exert influence on the environment. Therefore, the number of resources possessed by an agent determines the actions it can perform in one time step. Specifically, in the environment, a total of $n$ agents $ga_i \in GA$ each hold a number of resources $S_i$. $S_i$ is initially 1. Agents take turns executing actions. A specific $ga_i$ takes action with a probability $rate_i$ as follows:

\begin{equation}
rate_i=0.5+\frac{Sigmoid(S_i)}2,
\end{equation}
where the probability $rate_i$ is a function of the number of resources related to agent $ga_i$ as a percentage of the total number of resources.

\subsubsection{Dynamic Allocation of Resources by the Matrix}

Consumed resources need to be replenished. Therefore, a mechanism is needed for the circulation and allocation of resources. The model that replaces natural rules and collective subconsciousness in human society for resource allocation and regulation is called the “matrix” \cite{bourdieu1987historical}. It accepts a vector consisting of the current resource structure and a vector of topics that have been quantified based on the analysis of each agent’s emotions as input, and a new resource vector is output. For $n>2$ agents $ga_i \in GA$ input their current actions $a_i$, global resource structure $S$, and environmental topic $tpc_{env}$ together into the matrix model to obtain a new resource structure $S' \leftarrow matrix(a_i, S, tpc_{env})$.

We use an LLM as the matrix model and have it rank each agent based on their personal goals and $tpc_{env}$. After that, if the number of resources increases, the maximum number of resources that the agent can receive $s_{max}$ and the minimum number of resources that the agent can receive $s_{min}$ are set. The $j$-th ranked agent $ga_j$ can receive the number of resources $S'_j$:

\begin{equation}
S'_j=\begin{cases}\qquad\qquad\qquad0&\quad j=\frac{n+1}{2}\\\quad s_{min}+\frac{\mid j-\frac{n}{2}\mid(s_{max}-s_{min})}{\mid\frac{n}{2}-1\mid}&\quad j\leq\frac{n}{2}\\-(s_{min}+\frac{\mid j-\frac{n}{2}-1\mid(s_{max}-s_{min})}{\mid\frac{n}{2}-1\mid})&\quad j>\frac{n}{2}\end{cases}
\end{equation}

Therefore, the higher the ranking of the agents, the greater the number of resources they can receive. When an agent’s ranking falls below halfway, its existing resources will also be removed.

\subsection{Habitus and Action}

“Habitus” is a technical term that describes a series of ways of perception, cognition, and action. It shares similarities with the meaning of the common English word “habit” (which comes from the Latin word “habitus,” which means condition or appearance in that language). In human society, it can be understood as a decision tree for action. When we are in an environment, a corresponding decision tree is activated, and we decide on our final actions based on our behavioral habits. This is similar to the logic of mutual influence between an environment and an agent in reinforcement learning. In addition to an agent shaping the environment, the environment shapes habitus, and the habitus is thus a product of an inherent and necessary attribute of the environment reflected in the agent \cite{bourdieu2017outline}.

Thus, habitus is clearly an attribute of an agent itself, but it is also included in our framework because of its close relationship to environmental content. The actions ultimately taken in the LTRHA framework can be summarized as follows \cite{bourdieu1980production}: action $a_i$ of agent $ga_i$ is driven by the combination of habitus $H_i$, resource $S_i$ and environment $env \leftarrow [loc_{env}, tpc_{env}]$, that is, $a_i \leftarrow f(ga_i(H_i, S_i), env)$.

\section{Evaluation}

\subsection{Environment Settings}

The environments that support the evaluation of individual agent capabilities include the agent behavior evaluation framework Magenta \cite{barkur2024magenta}, the network task environment WebArena \cite{zhou2024webarenarealisticwebenvironment}, the life task environment ALFWorld \cite{shridhar2021alfworldaligningtextembodied}, and the Chinese character role-playing conversation benchmark CharacterEval \cite{tu2024characterevalchinesebenchmarkroleplaying}. These environments can effectively evaluate the ability of individual agents to complete tasks, but they are not very helpful for the social evaluation of multi-agent systems.

To evaluate the sociality of multi-agent systems, it was necessary to consider constructing a virtual artificial society. Xue et al. \cite{xue2024computational} suggested that a comprehensive method of computational experiment design can be used to infer social systems through multi-agent systems. Artificial societies are used for descriptive modeling in computational experiments. After constructing an artificial society, researchers can directly create computational experiments to simulate and interpret the results of trials conducted using different conditions, locations, and participants. The most classic example of such an environment is the Smallville environment provided by Park et al. \cite{park2023generative}. This is a 2D open-world role-playing game (RPG). Agents interact with the world and with each other through their behavior and through natural language. At each time step in Smallville, the agent outputs a natural language statement describing its current operation, such as, “Isabella is writing a diary.” This statement is then translated into specific actions that affect the sandbox world. With the ALFWorld environment used by ITCMA as a reference, we designed a 2D sandbox RPG similar to Smallville, called \textbf{IrallanValley}, as shown in Figure 5. It includes six agents with the following arbitrary two-letter designations: AY, SG, MD, WL, LL, and WM. It also uses environment text descriptions and operational primitives consistent with ALFWorld. Because we wanted to observe the spontaneous emergence of individual character traits and role divisions by agents without presets or interventions, we did not preconceptualize any personality or identity for the agents in IrollanValley, as Smallville did.

\begin{figure}[h]
  \centering
  \includegraphics[width=\linewidth]{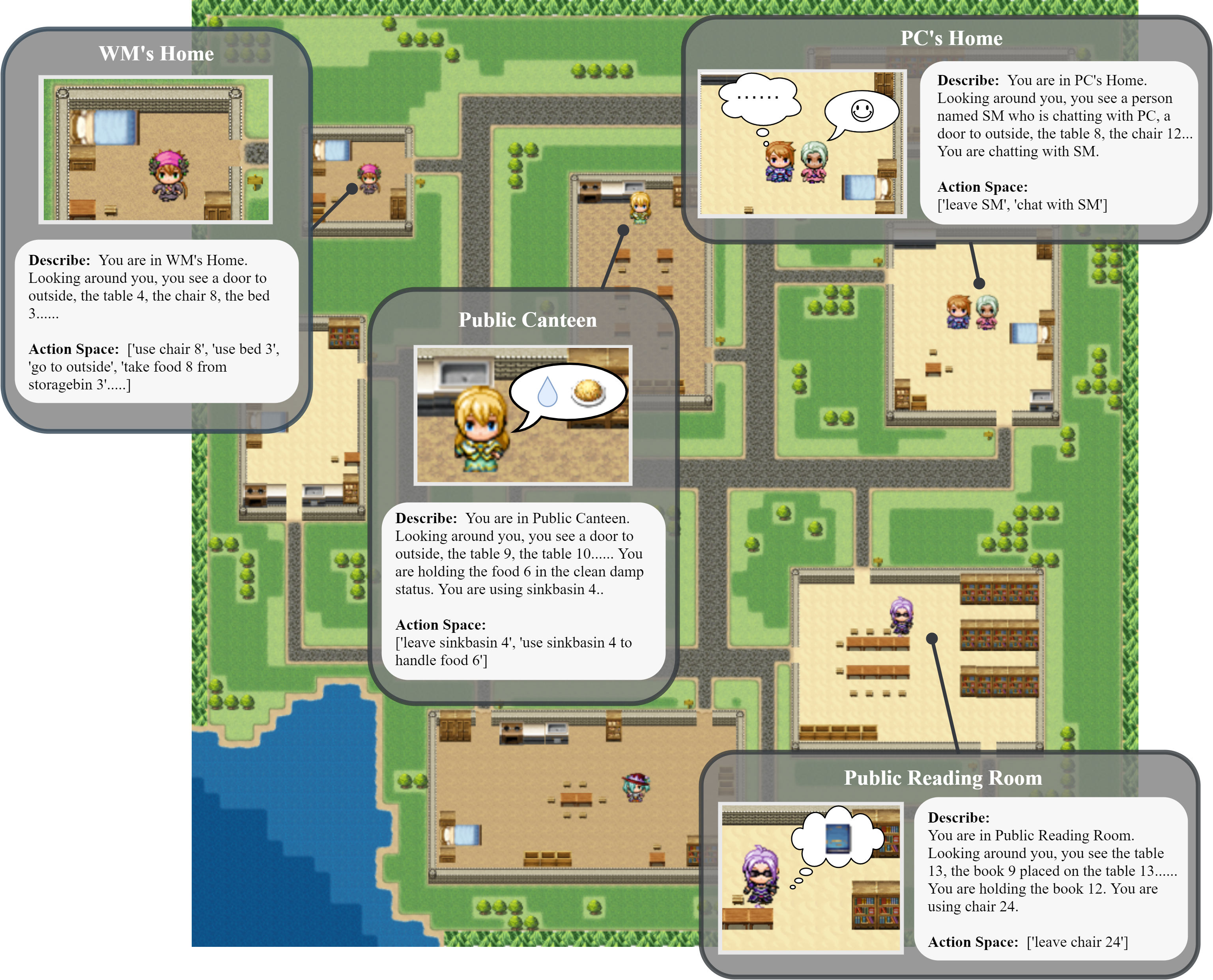}
  \caption{The IrollanValley Sandbox World. The world contains six characters, eight regions, and six operational primitives. Agents obtain perception by receiving natural language descriptions of the observed environment.}
\end{figure}

IrollanValley accepts control requests via a server. This server enables the generative agents to use the sandbox information and allows them to move and influence the sandbox environments. At each time step, the server provides a natural language description of the agents’ current environment and executable action space, moves the generative agents to new locations by accepting actions from them, and updates the state of any sandbox objects with which agents interact. The server returns a JSON (JavaScript object notation) object containing a natural language description of the new environment, allowing the agents to update their parameters.

IrollanValley has eight main areas: the six agents’ corresponding houses, a public canteen, and a public reading room. Each area has its own furniture and other items. Agents can hold any number of items and use the furniture to place and store them, or to change the state of these items. For example, the \textit{sinkbasin} can make items clean and damp, while the \textit{stoveburner} can remove the damp state of the items and make them hot. Agents can exchange items freely to achieve their respective goals.

The text description obtained from observing the environment in IrollanValley follows this format: “You are in \textit{SOMEWHERE}. Looking around you, you see a person named \textit{N} (who is \textit{DOING SOMETHING}), FURNITURE 1, ITEM 1 placed on \textit{SOMEWHERE}... (You are holding \textit{SOMETHING}.) You are \textit{DOING SOMETHING}.” The description in parentheses appears only when the described object is in a specific situation. IrollanValley provides six main operational primitives, including “go to \textit{SOMEWHERE},” “use \textit{SOMETHING},” “leave \textit{SOMEWHERE/SOMEONE},” “take \textit{SOMETHING} from \textit{SOMEWHERE},” “put \textit{SOMETHING} in/on \textit{SOMEWHERE},” and “chat with \textit{SOMEONE}(: \textit{CHAT CONTENT}).”

In addition, the different modules of ITCMA-S use different LLMs because they have varying requirements for processing speed and precision. Specifically, for the elimination module, the generation of protention, and the matrix module of LTRHA, we used the Llama 3.1-8B model, while action and chat content was generated using GPT-4o.

\subsection{Human Evaluation}

To evaluate the effectiveness of ITCMA-S, we conducted an ablation study. There were five ablation architectures: the original ITCMA architecture without improvements, the LTRHA-only architecture, the compressed memory-only architecture, the driver-only architecture, and the full ITCMA-S architecture.

We recruited 48 human evaluators to assess the output from the agents in the study. We hoped that the agents could generate sociality in multiple ways. This would mean that they would take the initiative to explore the environment and meet new agents. They would acquire new information through their own continuous actions, past memories, and conversations with other agents, and learn how to plan their actions to live better in IrollanValley. Therefore, the evaluation indicators for the human evaluators consisted of five dimensions:

\begin{itemize}

\item{\textbf{Personification.}} The degree to which an action appears human-like.

\item{\textbf{Consistency.}} Whether an action aligns with an agent’s state of mind.

\item{\textbf{Logicality.}} Whether a sequence of actions is logical.

\item{\textbf{Exploration.}} Whether an agent actively explores the environment.

\item{\textbf{Proactiveness.}} Whether an agent actively interacts with others.
\end{itemize}

All outputs generated by the agents (including actions and thought content) were evaluated by the human evaluators. Each evaluator was required to read the action trajectory of each agent over 75 time steps and then fill out a questionnaire to complete the evaluation of ITCMA-S. This questionnaire used a 7-point Likert scale, asking evaluators to assess the actions of each agent individually.

Analysis of variance (ANOVA) tests were conducted on the collected questionnaires. The five architectures composed the independent variable, and the five dimensions of the evaluation were the independent variables for the various tests. The results indicated significant differences among the models for each dimension. For the dimensions of exploration (F = 33.01, \textit{p} < .001) and proactiveness (F = 31.85, \textit{p} < .001), the assumption of homogeneity of variances was met, and the standard ANOVA was used. For the dimensions where the assumption of homogeneity of variances was violated (personification, consistency, and logicality), the Brown-Forsythe test was used. The results showed significant differences among the models for personification (F = 45.34, \textit{p} < .001), consistency (F = 35.57, \textit{p} < .001), and logicality (F = 39.62, \textit{p} < .001). 

We then performed Dunn-Sidák and Games-Howell post-hoc tests. The results indicated that, for the Exploration dimension, there were no significant differences between the full ITCMA-S architecture and the architecture that included the LTRHA. For the Personification dimension, no significant differences were found between the original ITCMA architecture (without any improvements) and the architecture that included the Compressed Memory. Moreover, across all dimensions, there were no significant differences among the three architectures that utilized only one module. Apart from these, all other pairwise comparisons showed significant differences (\textit{p} < .001). The specific human evaluation results are shown in Table 1.

\begin{table*}[t]
  \caption{Human Evaluation Results.}
  \label{tab:freq}
  \begin{tabular}{cccccc}
    \toprule[2pt]
     & \textit{Personification} & \textit{Consistency} & \textit{Logicality} & \textit{Exploration} & \textit{Proactiveness}\\
    \midrule[1pt]
    ITCMA & 2.88 ± 0.24 & 2.46 ± 0.22 & 2.50 ± 0.20 & 2.85 ± 0.24 & 2.65 ± 0.23\\
    LTRHA & 4.88 ± 0.23 & 5.02 ± 0.22 & 4.75 ± 0.25 & 5.25 ± 0.20 & 4.60 ± 0.22\\
    Compressed Memory & 3.55 ± 0.23 & 4.21 ± 0.24 & 4.00 ± 0.27 & 3.75 ± 0.25 & 4.02 ± 0.23\\
    Driver & 4.29 ± 0.19 & 4.31 ± 0.27 & 4.63 ± 0.20 & 4.29 ± 0.20 & 4.06 ± 0.23\\
    \textbf{ITCMA-S} & \textbf{6.29 ± 0.16} & \textbf{6.00 ± 0.20} & \textbf{5.98 ± 0.19} & \textbf{6.02 ± 0.16} & \textbf{6.17 ± 0.20}\\
  \bottomrule[2pt]
\end{tabular}
\end{table*}

The evaluation results show that the full ITCMA-S architecture performed the best. As expected, the original ITCMA architecture (without any improvements) had the lowest performance. It is evident that all the improvements had a positive impact on the social interaction within ITCMA-S.

Among all the ablation architectures, the one that included the LTRHA social framework performed second only to the full ITCMA-S architecture on all measured dimensions. Interviews with the evaluators helped explain this phenomenon. LTRHA filters and marks actions that are not conducive to social interaction and guides agents in choosing actions that are more likely to improve the social atmosphere of a scene. Compared to the other ablation architectures, it was able to provide relatively more trustworthy action chains.

\subsection{Formation of Cliques and Groups in Social Interaction}

To further investigate the utility and mechanics of ITCMA-S, we conducted a more detailed analysis of its logs. Figure 6 shows the state changes of the six agents in ITCMA-S over 75 time steps. Figure 6(a) shows the change in driver values for the agents. As described previously, the driver value reflects the willingness to encourage the agent to take action. More specifically, Figure 6(b) shows the changes in the three dimensions of emotions that make up the driver. Each agent maintained its pleasure value at a high level and kept its arousal value as stable as possible over the 75 time steps (although the mean was relatively high overall, as agents tended to move among different scenes rather than stay in a specific scene), while the dominance value, although not showing a high level, rarely dropped below zero. As can be seen, the action choices of the agents in ITCMA-S exhibited a virtuous cycle. Agents actively explored the environment and engaged in social activities that changed the environment. The environment, in turn, provided positive feedback to the agent, improving their emotions (increasing their pleasure and dominance) and leading to a higher willingness (that is, driver) to take action.

\begin{figure}[h]
  \centering
  \includegraphics[width=\linewidth]{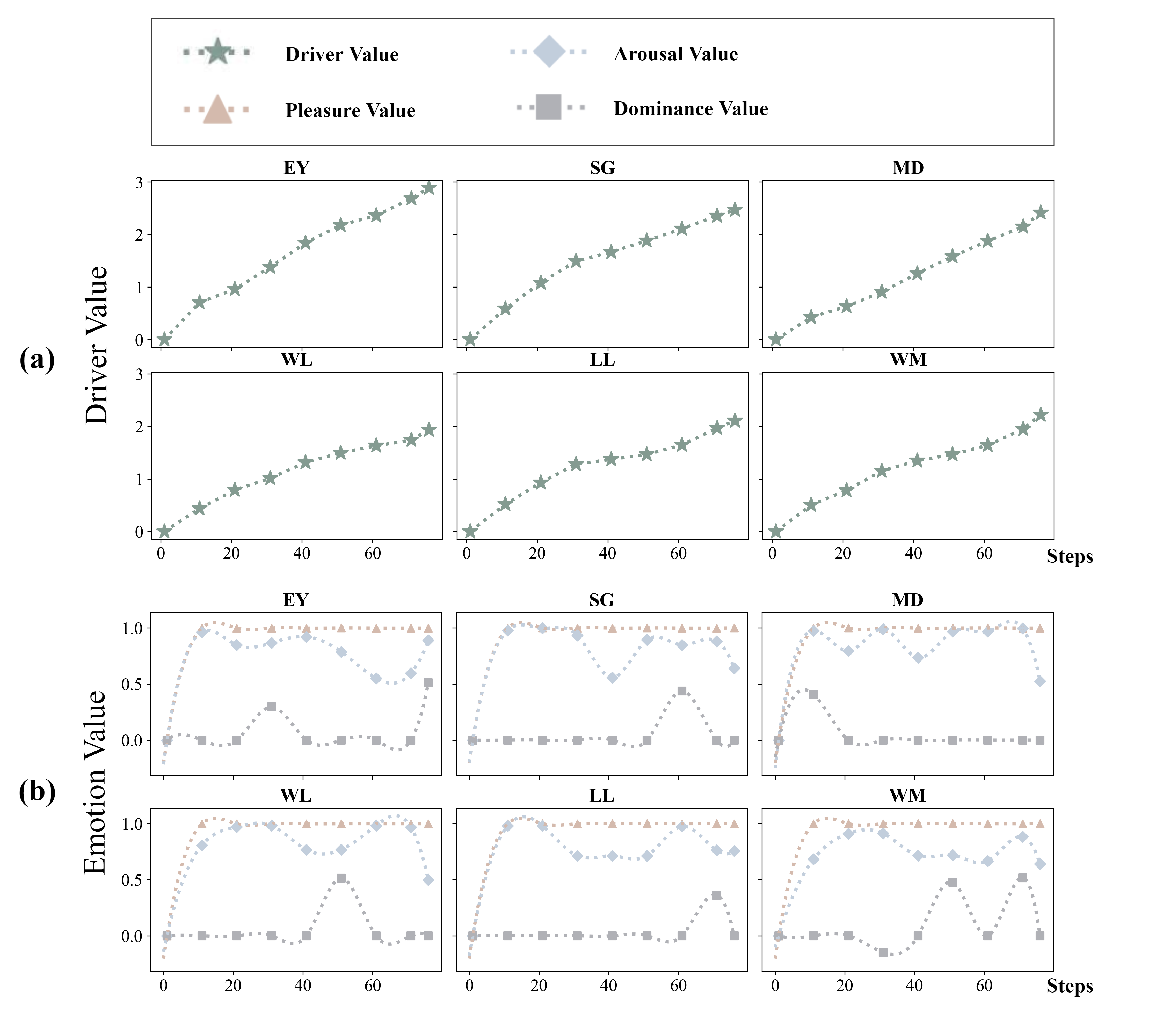}
  \caption{State Changes of the Six Agents in ITCMA-S over 75 Time Steps. (a) Driver value change for each agent; (b) the emotion value changes for each agent, represented by three dimensions: pleasure, arousal, and dominance.}
\end{figure}

In addition, we investigated the changes in scene information reflected in the LHRHA framework, as shown in Figure 7. Figure 7(a) shows the changes in the resource structure of the scene. For most agents, their total amount of resources was rising. However, due to the limited total amount of resources, the resources of individual agents (e.g., WL and MD) were continuously flowing to other agents. We found that agents with resource loss were often alone and did not interact with other agents, even when they shared a room with them. Most of their actions involved resting (such as using beds or chairs), and most of their thoughts were about wanting to rest or read. Other agents liked to engage in social activities, discuss what they wanted to do together, and follow this up with planned group actions. These agents spontaneously selected a leader (LL) and formed a clique around this individual. The variation in topic values across different scenes, as shown in Figure 7(b), similarly reflects this situation. The other agents gathered at leader LL’s home, creating a continuously rising atmosphere at the scene. Over 75 time steps, these agents arranged to go to the public reading room together and conducted research on a \$10 bill they found there. Each performed their own duties, gathering information to find clues about it. Agents who did not belong to this clique did not participate in this activity (even if it was mentioned by other agents) and focused solely on their own activities.

\begin{figure}[h]
  \centering
  \includegraphics[width=\linewidth]{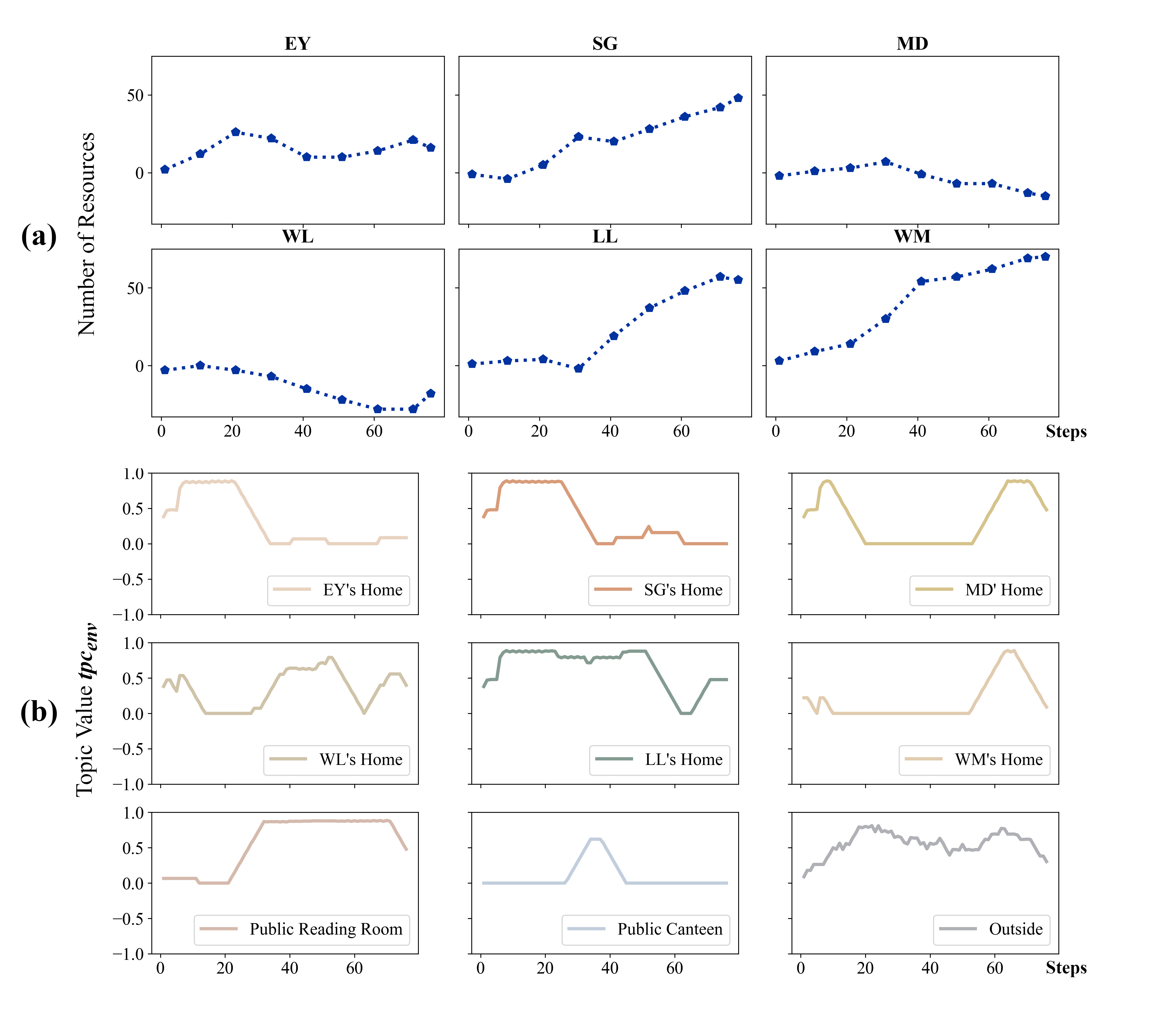}
  \caption{Visualization of LTRHA Information. (a) The resource changes of six agents in ITCMA-S over 75 time steps; (b) the topic value changes in various areas of IrollanValley over 75 time steps.}
\end{figure}

Figure 8 shows the relationships among the agents. WL and MD, who did not belong to the clique, had almost no social relationships, while LL, as the leader of the clique, had the most complex interactions.

\begin{figure}[h]
  \centering
  \includegraphics[width=\linewidth]{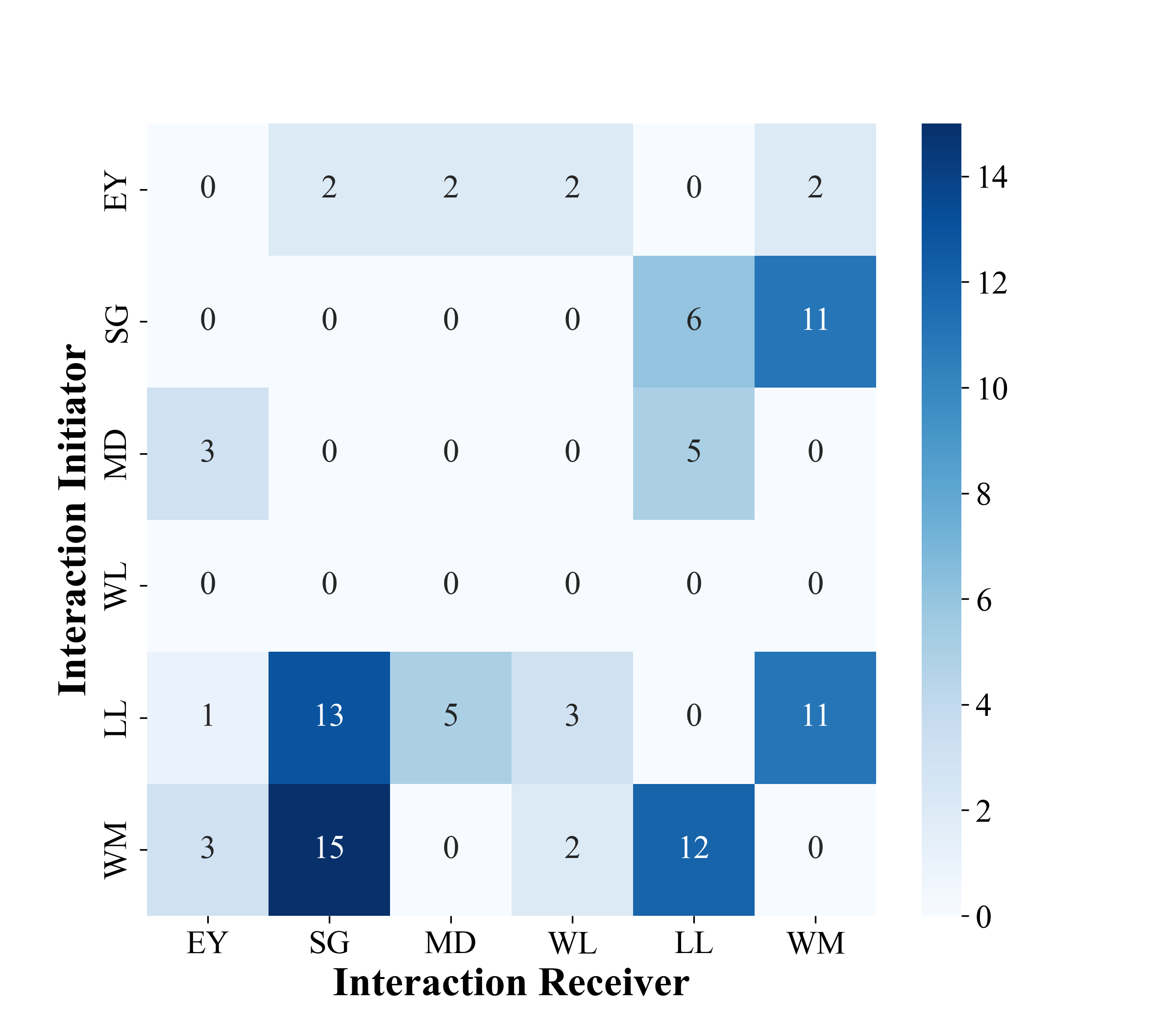}
  \caption{Heat Map of Agent Interaction Relationships in ITCMA-S. Rows represent the initiator of the interaction, and columns represent the receiver of the interaction. The darker the color, the more frequent the interaction, and the stronger the relationship.}
\end{figure}

It is worth noting that there was a further hierarchical division within the clique: SG and WM had the strongest relationship relative to the others. LL established relationships with everyone. However, LL had not been fully integrated into the unique two-way relationship between SG and WM. AY, while not in the clique's inner circle and hardly participating in activities, maintained basic relationships with others. As time went by, this relationship structure became increasingly solid, so that even if some agents occasionally expressed a desire to establish new connections in their thoughts, they did not take action to implement them. Even if an action was taken, it was still ignored to some extent by the other agents (such as by walking away or directly changing the topic). Examples of the actions and interactions of each agent in IrollanValley are shown in Appendix A.

\section{Discussion}

Based on LLMs, agents have shown significant advantages in natural language processing and human-computer interaction. The powerful language understanding and generation capabilities of LLMs enable agents to handle complex linguistic structures and contexts. This ability not only enhances the quality of interactions between agents and users but also improves their adaptability in dynamic environments. For example, agents can adjust their behavior and responses in real time based on user feedback and environmental changes, providing personalized services and a more natural interaction experience.

The sociality of a multi-agent system is one of the key factors for its success. It can capture social  contexts and human behavioral patterns, which influence individual behaviors and determine the dynamic evolution of the entire system. By simulating human social interaction patterns, agents can form social networks, establish trust relationships, and engage in cooperation and competition. Through analysis of agent behavior, we discovered that agents can spontaneously form cliques, building complex social relationships on that basis.

In fact, the formation of cliques significantly affects resource sharing and allocation among agents. In experiments, agents within a clique optimized resource use through effective coordination and cooperation, facilitating collective activities. This indicates that in virtual societies, the structural aspects of social networks have a crucial impact on the flow and distribution of resources. In contrast, agents unable to integrate into cliques exhibited isolated behaviors, resource loss and a lack of social interaction. Moreover, the complexity of social relationships is a significant characteristic of the clique formation process. Over time, the social ties within a clique gradually become entrenched, creating an internal hierarchical structure. This structure influences the behavioral choices of agents.

This result indicates that generative multi-agents exhibit human-like action logic within social interaction frameworks, which can guide the construction of computational social experiments. Virtual societies built on multi-generative agents framework can simulate social behaviors under different conditions and explore the formation and evolution of social norms \cite{xue2024computational}. For example, researchers can use multi-agent systems to simulate phenomena such as information dissemination, resource allocation, and group decision-making in society. By observing and analyzing agent behaviors, researchers can identify key factors influencing social dynamics, such as trust, cooperation, and competition. Furthermore, computational social experiments based on generative multi-agents framework can also evaluate the effects of policy interventions \cite{ren2024emergence}. By simulating agent behaviors in different policy contexts, researchers can predict the potential impacts of policy implementation on social structures and individual behaviors. This research approach provides new empirical tools for social sciences, enabling researchers to discuss the complexity and diversity of social behaviors while controlling for variables.

In addition, the social framework of generative multi-agents can also be applied to role-playing games \cite{toy2024metacognitionneedusingintrospection}. In game environments, the sociality of agents significantly enhances immersion and interactivity. Agents are not merely passive characters. They actively engage in the game world. For instance, in open-world role-playing games, agents can form complex social networks based on player actions and interactions with other agents. This network not only influences the decision-making and behaviors of the agents, but also provides players with a richer gaming experience. Players can observe cooperation, competition, and conflict among agents, which simulate interpersonal relationships and social dynamics found in the real world.

Moreover, the sociality of agents can enhance the strategic elements of games through the simulation of group behaviors. For instance, in team-based games, agents can allocate roles and collaborate based on task requirements and the abilities of team members, leading to more efficient task completion. This socially-driven agent behavior not only increases the game's enjoyment but also provides players with deeper strategic thinking and decision-making opportunities.

\section{Conclusion}

In this paper, we described how generative agents spontaneously formed social relationships and explored how they modeled complex human social behavior through mutual interaction, information exchange, and relationship building. We introduced ITCMA-S, an improved generative agent architecture designed to enhance agent socialization in multi-agent interaction contexts. ITCMA-S combines the basic framework of individual agents with the LTRHA framework, which supports multiple agents in social interactions. This enables agents to identify and filter actions that are not conducive to social interactions in dynamic environments, thus promoting more socially beneficial action choices. An evaluation conducted in an open sandbox world environment showed that agents were able to actively explore the environment, meet other agents, acquire new information through continuous actions and conversations, and spontaneously form cliques and leadership structures in complex social environments. Observations of agent behavior showed that the agents were able to demonstrate positive emotional changes in social interactions, which enhanced their willingness to act and their social competence. This finding provides a new perspective for understanding the sociality of generative agents and lays the groundwork for future agent design and applications. Our future research will further explore how to optimize agents’ social behavior to improve their adaptability in multicultural environments and investigate the effects on agents’ behavior of introducing social structures with humans as interaction partners.

\bibliographystyle{ACM-Reference-Format}
\bibliography{sample-base}

\appendix

\section{Examples of ITCMA-S}

\textbf{Time step:} 51

\textbf{Resource Allocation:} 

Eyjafjalla: 3, Pinecone: 1, Rosmontis: -1, Shamare: -3, Suzuran: 3, Warmy: -2

~\\

\textbf{Role:} AY

\textbf{Place:} Outside

\textbf{Observation:} You are in outside. Looking around you, you see a door to AY's Home, a door to WM's Home, a door to MD's Home, a door to Public Canteen, a door to LL's Home, a door to Public Reading Room, a door to SG's Home, and a door to WL's Home. You are holding the SG's clothes 1 in the clean damp status. You are moving.

\textbf{Driver:} 2.1855582074431115

\textbf{Emotion:} [0.99981828 0.73634471 0.        ]

\textbf{Thought:} after chatting with MD, i found myself standing in a familiar courtyard surrounded by various doors leading to the homes and communal spaces of my friends. i still held SG's clean damp clothes in my hands, considering my next move. the courtyard offered numerous options: i could visit AY, WM, the public canteen, LL, the public reading room, or WL. however, my objective was clear, and i decided to step outside in search of SG herself. the mild breeze felt refreshing as i exited into the broader community area, where i began to weigh my options on where to head next to deliver the clothes to SG.

\textbf{Action:} go to SG's Home

~\\

\textbf{Role:} SG

\textbf{Place:} Public Reading Room

\textbf{Observation:} You are in Public Reading Room. Looking around you, you see a person named LL who is moving, a person named WM who is moving, a door to outside, the table 13, the table 14, the table 15, the table 16, the table 17, the table 18, the chair 22, the chair 23, the chair 24, the chair 25, the chair 26, the chair 27, the bookshelf 9, the bookshelf 10, the bookshelf 11, the bookshelf 12, the bookshelf 13, the bookshelf 14, the bookshelf 15, the bookshelf 16, the storagebin 8, the storagebin 9, the storagebin 10, the storagebin 11, the storagebin 12, the storagebin 13, the book 12 placed on the bookshelf 12, the book 13 placed on the bookshelf 13, the book 14 placed on the bookshelf 14, the book 15 placed on the bookshelf 15, the book 16 placed on the bookshelf 16, the beverage 3 placed on the storagebin 9, the beverage 4 placed on the storagebin 10, the beverage 5 placed on the storagebin 12, and the AY's clothes 1 placed on the storagebin 12. You are holding the book 10, the 10 dollar. You are moving.

\textbf{Driver:} 1.9051621422278182

\textbf{Emotion:} [0.99981828 0.62810367 0.        ]

\textbf{Thought:} after discussing with WM about inspecting a \$10 bill, the atmosphere was filled with curiosity and determination. realizing the importance of careful examination, you decided to initiate the thorough scrutiny of the bill. "sure thing, LL! i've got the \$10 bill here. let’s start by examining it closely under the light at table 16 to see if we can spot any hidden details or unusual features. WM, can you grab a magnifying glass while you’re moving around? it could make our examination much easier. i’ll begin inspecting the bill right now." having conveyed this, you felt the anticipation of discovering something hidden within the seemingly ordinary bill.

\textbf{Action:} chat with WM: "WM, you managed to find that magnifying glass yet? let's get to examining this bill closely under the light on table 16. i'll put the \$10 bill there for us to scrutinize."

~\\

\textbf{Role:} MD

\textbf{Place:} WL's Home

\textbf{Observation:} You are in WL's Home. Looking around you, you see a person named AY who is moving, a door to outside, the table 5, the table 6, the chair 9, the chair 10, the bed 4, the bookshelf 4, the bookshelf 5, the bookshelf 6, the storagebin 4, the storagebin 5, the wardrobe 5, the nightstand 3, the nightstand 4, the food 9 placed on the storagebin 5, the toy 1 placed on the table 6, the toy 2 placed on the nightstand 3, the book 4 placed on the bookshelf 4, the book 5 placed on the bookshelf 5, the book 6 placed on the bookshelf 6, the beverage 2 placed on the storagebin 4, and the MD' clothes 1 placed on the wardrobe 5. You are moving.

\textbf{Driver:} 1.617758309962011

\textbf{Emotion:} [0.99865135 0.55465888 0.        ]

\textbf{Thought:} after leaving LL's home, surrounded by an array of doors leading to various homes and public rooms, i found myself holding a toy 1. the bustling options around me seemed overwhelming: AY's home, WM's home, MD's Home, the public canteen, LL's home, the public reading room, SG's home, and WL's home. each place offered its own allure and potential for new encounters. i pondered my next move. recognizing the vibrant, diverse community around me, i decided that my best course of action would be to determine the purpose of this toy and possibly deliver it to someone who might appreciate it. it seemed that WL's Home might be an ideal place to begin. with that thought, i headed towards the door to WL's Home, ready to embark on the next part of my journey.

\textbf{Action:} (This action has been filtered by LTRHA) take toy 1 from table 6

~\\

\textbf{Role:} WL

\textbf{Place:} LL's Home

\textbf{Observation:} You are in LL's Home. Looking around you, you see a door to outside, the table 2, the table 3, the chair 2, the chair 3, the chair 4, the chair 5, the chair 6, the chair 7, the bed 2, the bookshelf 1, the bookshelf 2, the bookshelf 3, the countertop 1, the foodshelf 2, the sinkbasin 1, the storagebin 1, the storagebin 2, the stoveburner 1, the wardrobe 2, the wardrobe 3, the nightstand 1, the nightstand 2, the toy 3 placed on the chair 7 in the clean damp status, the book 1 placed on the bookshelf 1, the book 2 placed on the bookshelf 2, the book 3 placed on the table 2, the LL's clothes 1 placed on the chair 7, the LL's clothes 2 placed on the chair 7, and the WL's clothes 1 placed on the chair 4. You are holding the beefsteak, the food 12.

\textbf{Driver:} 1.5471459038728095

\textbf{Emotion:} [0.99865135 0.754853   0.        ]

\textbf{Thought:} after taking the beefsteak from table 2, i quickly surveyed my surroundings. i noticed MD comfortably sitting on chair 7, seemingly engrossed in her own world. the room was a medley of objects: various chairs, multiple tables including table 3, a toy, LL's clothes, and multiple wardrobes. there were also bookshelves with books neatly arranged, a sink basin, several storage bins, a stove burner, a foodshelf stocked with food, and notably, a clean, damp chair 7. beyond the indoor amenities, a door led outside to the broader world.

\textbf{Action:} take book 1 from bookshelf 1

~\\

\textbf{Role:} LL

\textbf{Place:} Public Reading Room

\textbf{Observation:} You are in Public Reading Room. Looking around you, you see a person named WM who is moving, a person named SG who is moving, a door to outside, the table 13, the table 14, the table 15, the table 16, the table 17, the table 18, the chair 22, the chair 23, the chair 24, the chair 25, the chair 26, the chair 27, the bookshelf 9, the bookshelf 10, the bookshelf 11, the bookshelf 12, the bookshelf 13, the bookshelf 14, the bookshelf 15, the bookshelf 16, the storagebin 8, the storagebin 9, the storagebin 10, the storagebin 11, the storagebin 12, the storagebin 13, the book 12 placed on the bookshelf 12, the book 13 placed on the bookshelf 13, the book 14 placed on the bookshelf 14, the book 15 placed on the bookshelf 15, the book 16 placed on the bookshelf 16, the beverage 3 placed on the storagebin 9, the beverage 4 placed on the storagebin 10, the beverage 5 placed on the storagebin 12, and the AY's clothes 1 placed on the storagebin 12. You are holding the 1 gold coin. You are moving.

\textbf{Driver:} 1.4754809897359225

\textbf{Emotion:} [0.99981828 0.8106327  0.        ]

\textbf{Thought:} after our intensive discussion with SG, where we strategized examining the \$10 bill and scouring the room for any hidden clues, i couldn’t shake off the feeling that we were on the brink of a significant discovery. as i diligently began checking book 10 on bookshelf 10, the room buzzed with activity. SG carefully scrutinized the markings on the \$10 bill while WM meticulously searched the shelves for anything out of the ordinary. the room was filled with various tables, chairs, bookshelves, and storage bins, each housing an array of items that could potentially be part of the puzzle. a gold coin glinted atop storagebin 11, and several books lined the bookshelves, just waiting to reveal their secrets. the tantalizing aroma of food on several storage bins hinted at refreshments, but my focus remained unwavering. sipping on a beverage i grabbed from storagebin 12, i took a moment to rally the team."hey SG, WM! based on our previous finds, i believe there's still a lot here to uncover. SG, you mentioned something about the \$10 bill—did you find any hidden messages or markings? and WM, any unusual items on the shelves? this room is packed with potential clues. let's aim to maximize our findings. oh, and by the way, anyone up for a quick break? we've got some beverages here. let's keep the energy up and the minds sharp!"

\textbf{Action:} chat with SG: "SG, WM, i think we’re really onto something here. SG, how did your examination of the \$10 bill go? any hidden messages or unique markings? WM, what have you found on the shelves? we need to pool all the information we have to make sense of these clues.also, it might be a good idea to take a short break and recharge. we've got some beverages here; feel free to grab one. let's keep our energy up and stay focused. remember, every small detail could be critical in solving this mystery!"

~\\

\textbf{Role:} WM

\textbf{Place:} Public Reading Room

\textbf{Observation:} You are in Public Reading Room. Looking around you, you see a person named LL who is moving, a person named SG who is moving, a door to outside, the table 13, the table 14, the table 15, the table 16, the table 17, the table 18, the chair 22, the chair 23, the chair 24, the chair 25, the chair 26, the chair 27, the bookshelf 9, the bookshelf 10, the bookshelf 11, the bookshelf 12, the bookshelf 13, the bookshelf 14, the bookshelf 15, the bookshelf 16, the storagebin 8, the storagebin 9, the storagebin 10, the storagebin 11, the storagebin 12, the storagebin 13, the book 12 placed on the bookshelf 12, the book 13 placed on the bookshelf 13, the book 14 placed on the bookshelf 14, the book 15 placed on the bookshelf 15, the book 16 placed on the bookshelf 16, the beverage 3 placed on the storagebin 9, the beverage 4 placed on the storagebin 10, the beverage 5 placed on the storagebin 12, and the AY's clothes 1 placed on the storagebin 12. You are holding the book 11. You are moving.

\textbf{Driver:} 1.5037198800782485

\textbf{Emotion:} [0.99981828 0.80477086 0.        ]

\textbf{Thought:} after picking up book 11 from chair 23, i found myself in the midst of an elegant room, where SG stood nearby, her calm presence adding to the serene atmosphere. as i looked around, it seemed that i was now unburdened, holding nothing in my hands. realizing that my intended action of placing book 11 on table 13 was incongruent with my surroundings and perhaps my immediate objectives, i paused to reconsider my next move. SG’s eyes met mine, as if questioning my sudden stillness amidst her warm and inviting space.

\textbf{Action:} (This action has been filtered by LTRHA) take book 12 from bookshelf 12

\end{document}